\documentclass[final,1p,times]{elsarticle} 
\usepackage{graphicx} 
\usepackage{amssymb} 
\usepackage{amsthm} 
\usepackage{lineno} 

\journal{Nuclear Physics A} 
\begin{document} 

\begin{frontmatter} 


\title{NA60 results on $\phi$ production in the hadronic and leptonic channels in In-In collisions at 158 GeV}

\author{Alessandro De Falco$^{a}$ for the NA60 collaboration}

\address[a]{Universit\`a and INFN Cagliari, 
Complesso Universitario di Monserrato,
09042 Monserrato (CA), Italy}

\begin{abstract} 
The NA60 experiment at the CERN SPS studied $\phi$ meson production in
In-In collisions at 158 A GeV via muon and kaon decay channels. Results in the 
hadronic channel are presented for the first
time. These are discussed in the framework of the so-called $\phi$
puzzle through the comparison with the previous NA60 measurements in
the muon channel. The yield and inverse $m_T$ slopes observed in the two
channels are compatible within errors, showing that the large
discrepancies seen in Pb-Pb collisions between NA50 (muon pairs) and 
NA49 (kaon pairs) are not seen in the NA60 In-In data. \\
\end{abstract} 

\end{frontmatter} 


Strangeness enhancement was proposed already in 1982 as a signature 
of the occurrence of the quark gluon plasma phase~\cite{Rafelski}. 
Due to its $s \bar s$ valence quark content, the $\phi$ meson is a particularly 
valuable probe for the measurement of strangeness production. 

$\phi$ production measurements were performed at the CERN SPS 
by the NA49, NA50 and CERES experiments in lead induced collisions
at 158~$A\cdot$GeV. Discrepancies were observed in the measurements 
performed by NA50~\cite{na50phi} in the muon channel and by 
NA49~\cite{na49phi} in the kaon channel, both
in Pb-Pb collisions. The $\phi$ multiplicity measured by NA50 is 
much higher than the one obtained by NA49.
Concerning the transverse mass distributions,  NA50 finds an inverse slope 
of about 230~MeV, almost independent of centrality, while
NA49 measures values increasing with the number of participants $N_{part}$ 
from $\sim$250~MeV to $\sim$300~MeV. 

The origin of this discrepancy, known as the $\phi$ puzzle,
has long been discussed. It was suggested  
that kaons may suffer rescattering and absorption in the medium, resulting in 
a depletion of kaon pairs, especially at low $p_T$, that leads to a reduced
yield and a hardening of the $p_T$ distribution in the hadronic channel. In this 
frame, the yield in lepton pairs is expected to exceed the one in kaon pairs  
by about $50\%$~\cite{kaondepl}.
 
Recent results by CERES~\cite{ceresphi} on $\phi$ production in 
Pb-Au collisions via the $K^+K^-$ and $e^+e^-$ decay channels 
confirm the NA49 results, and thus suggest that there is no
$\phi$ puzzle. However, the results in dielectrons, being affected by 
large statistical errors, are not conclusive. 
 
The NA60 experiment at the CERN SPS studied $\phi$ meson production in
In-In collisions both in muon and kaon pairs. Details of the apparatus are
reported in~\cite{usai}. 

The sample used for the results presented in this paper was collected with
a 158~A$\cdot$GeV In beam impinging on a segmented In target, composed
of 7 subtargets of 17\% total interaction probability placed in vacuum. 230 million 
triggers, mainly dimuons, were acquired. In order to avoid events with reinteractions of 
nuclear fragments in the subsequent targets, only events with one vertex
in the target region were selected. 

Results for dimuons, already presented in~\cite{epjphi}, 
are briefly summarized in the following. 
The dimuon sample consist of 440000 signal muon pairs.
The $\phi$ yield in the muon channel is measured counting the events in 
the mass window $0.98<M_{\mu \mu}<1.06$~GeV/$c^2$. 
The signal/background ratio below the $\phi$ peak, integrated over centrality,
is $\sim$1/3. 
To account for the continuum under the 
$\phi$ peak, two side windows between $0.88<M_{\mu \mu}<0.92$~GeV/$c^2$ 
and $1.12<M_{\mu \mu}<1.16$~GeV/$c^2$ are subtracted from the 
$\phi$ window. The raw transverse momentum distributions are extracted 
counting the number of $\phi$ mesons in several $p_T$ intervals.
They are then corrected for the acceptance using an overlay Monte Carlo 
tuned iteratively such that the resulting distributions reproduce the measured
data. 
The $\phi$ multiplicity was obtained either with a direct measurement of the cross 
section or using the $J/\psi$ as a calibration process. 
The two methods agree within $\sim10\%$. 

The analysis in the $KK$ channel is performed assuming that all the charged 
particles associated with the primary vertex are kaons, and building all the 
possible pairs among the tracks of each event. This results in a considerable 
combinatorial background, ranging from $\sim 170$ to $\sim 400$ times the signal, 
depending on centrality. The background has to be reduced with an appropriate set
of cuts and subtracted  with an event mixing technique. For the latter, events 
from runs taken in homogeneous conditions are grouped in 
pools according to the position of the target associated to the vertex, the 
centrality of the collision and the direction of the event plane. 
Tracks from different events belonging to the same pool are mixed.\\ 
Several cuts are applied both to the real and mixed samples. 
The vertices are required to be within 
$2\sigma_z$ from the center of the closest target, $\sigma_z \sim200~\mu$m 
being the resolution of the position of the interaction vertex in the longitudinal
coordinate. 
Events with a charged track multiplicity lower than 10 are discarded. 
The tracks associated with the vertex are selected requiring that the $\chi^2$ 
of the track fit is lower than 3. A cut on the track rapidity  
$2.9<y<3.7$  is applied. Additional cuts on the 
pairs are applied requiring that the opening angle $\theta_{KK}$ \
is within the range $0.005<\theta_{KK}<0.15$~rad and the transverse momentum is 
$p_T>0.9$~GeV/$c$. 
The obtained mixed spectra are normalized asking that in the mass region 
$1.02<m<1.06$~GeV/$c^2$ the number of mixed like-sign pairs coincides with 
the corresponding one in the real data.
%
%
\begin{figure}[ht]
\begin{minipage}[t]{0.47\textwidth}
\centering
\includegraphics[scale=0.30]{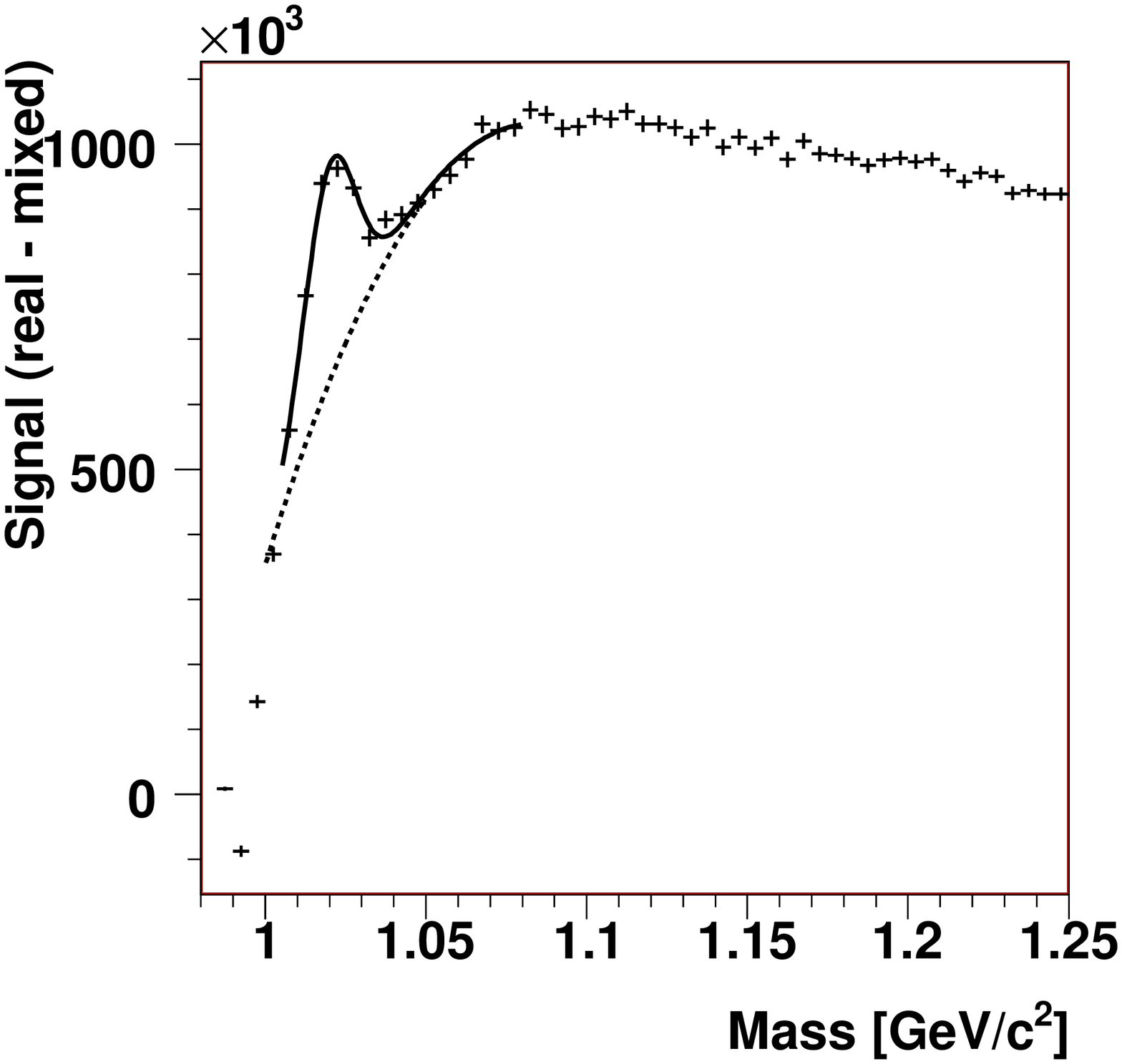}	       
\caption[]{Invariant mass spectrum of the opposite sign kaon pairs integrated in centrality.}
\label{fig:mass}
\end{minipage}
\hfill{ 
\begin{minipage}[t]{0.47\textwidth}
\centering
\includegraphics[scale=0.30]{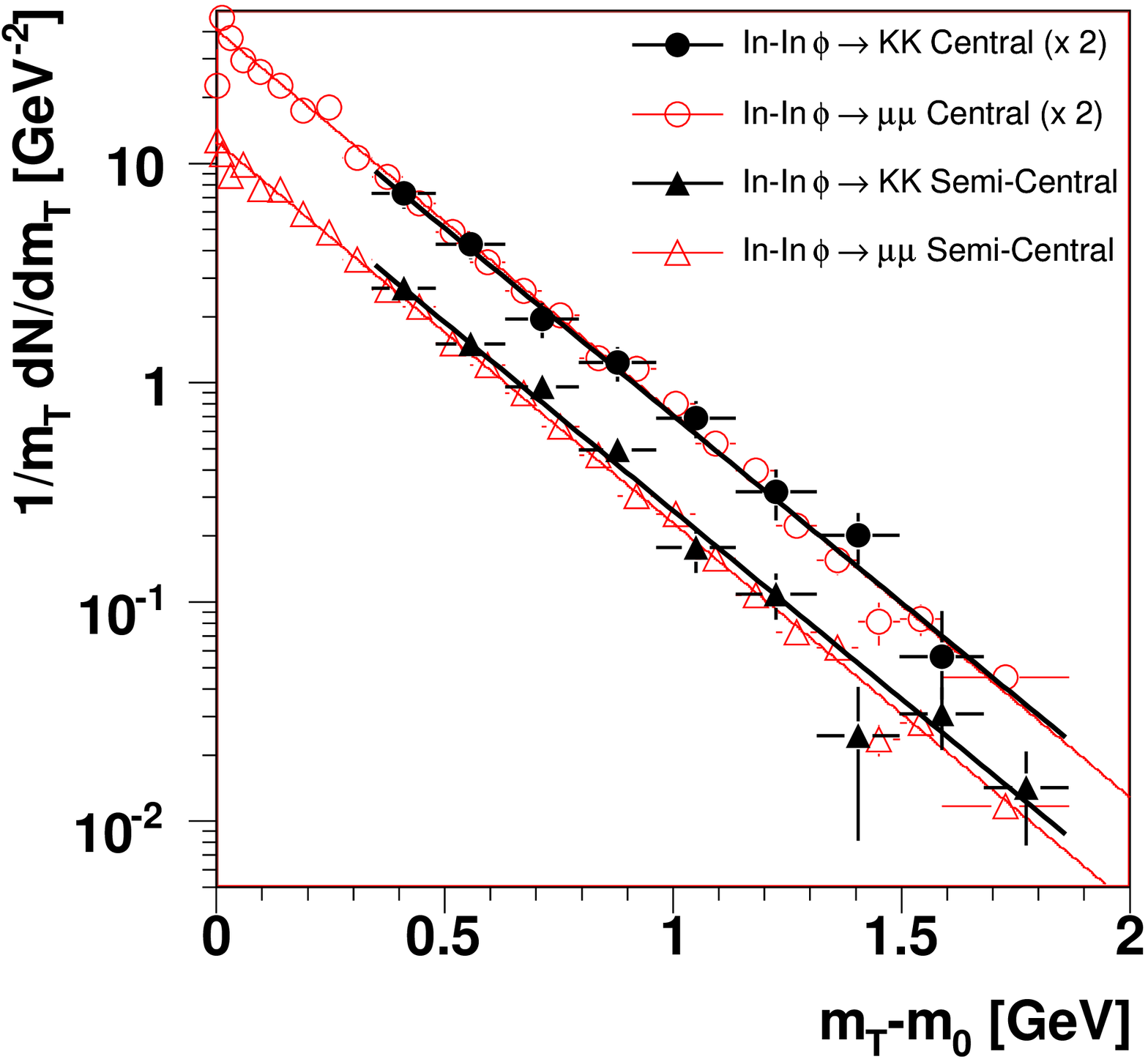}
\caption[]{$m_T$ spectra for semi-central (triangles) and central (circles)
collisions for $\phi\rightarrow KK$ (full symbols) 
and $\phi \rightarrow \mu\mu$ (open symbols).}
\label{fig:mtspectra}
\end{minipage}
}
\end{figure}
%
%

\begin{figure}[ht]
\begin{minipage}[t]{0.47\textwidth}
\centering
\includegraphics*[scale=0.33]{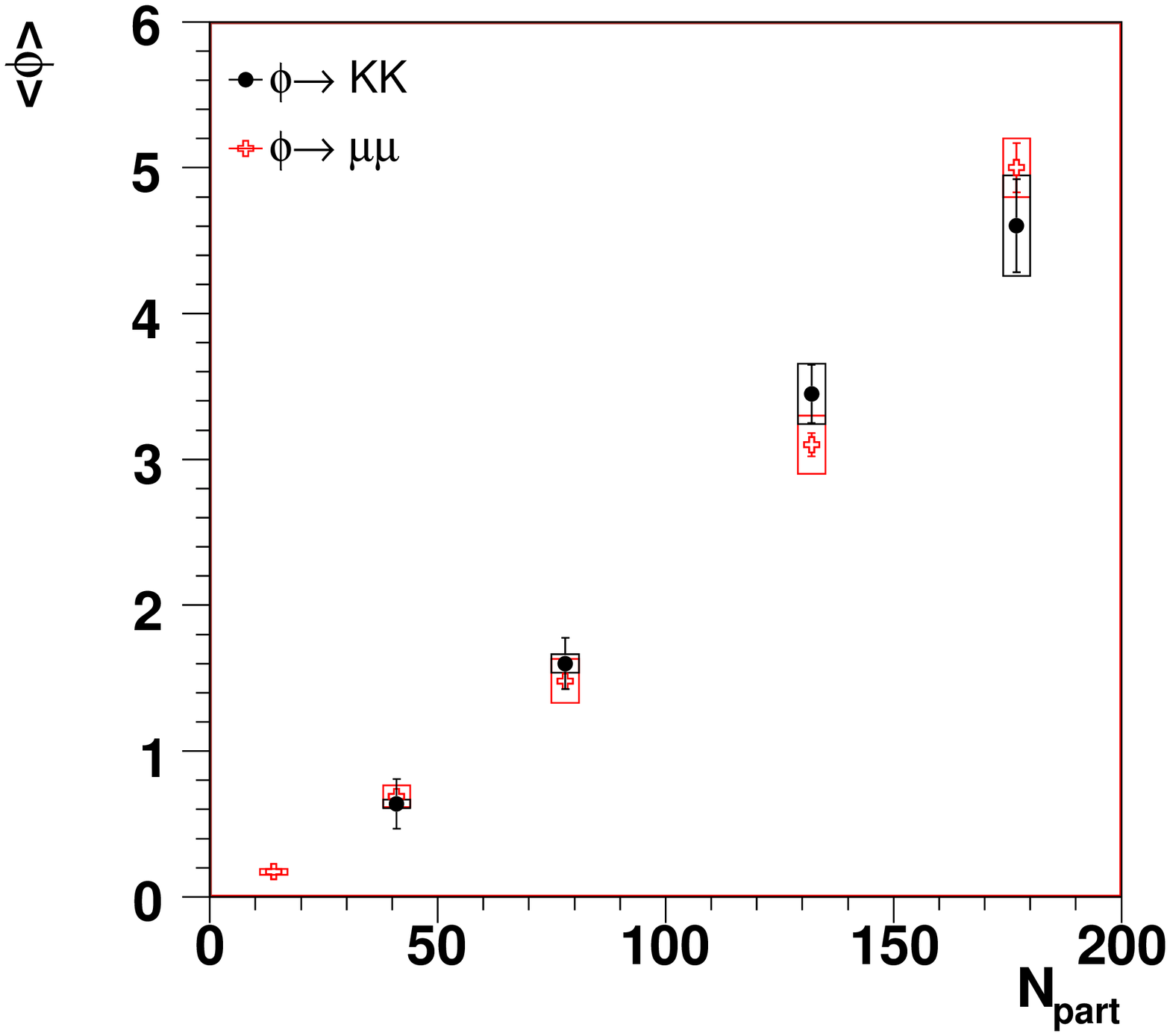}
\caption[]{Yield as a function of $N_{part}$ in In-In collisions in the 
$\phi\rightarrow KK$ (full circles) and $\phi\rightarrow \mu\mu$ 
channels (open crosses).}
\label{fig:yield}
\end{minipage}
\hfill{ 
\begin{minipage}[t]{0.47\textwidth}
\centering
\includegraphics[scale=0.31]{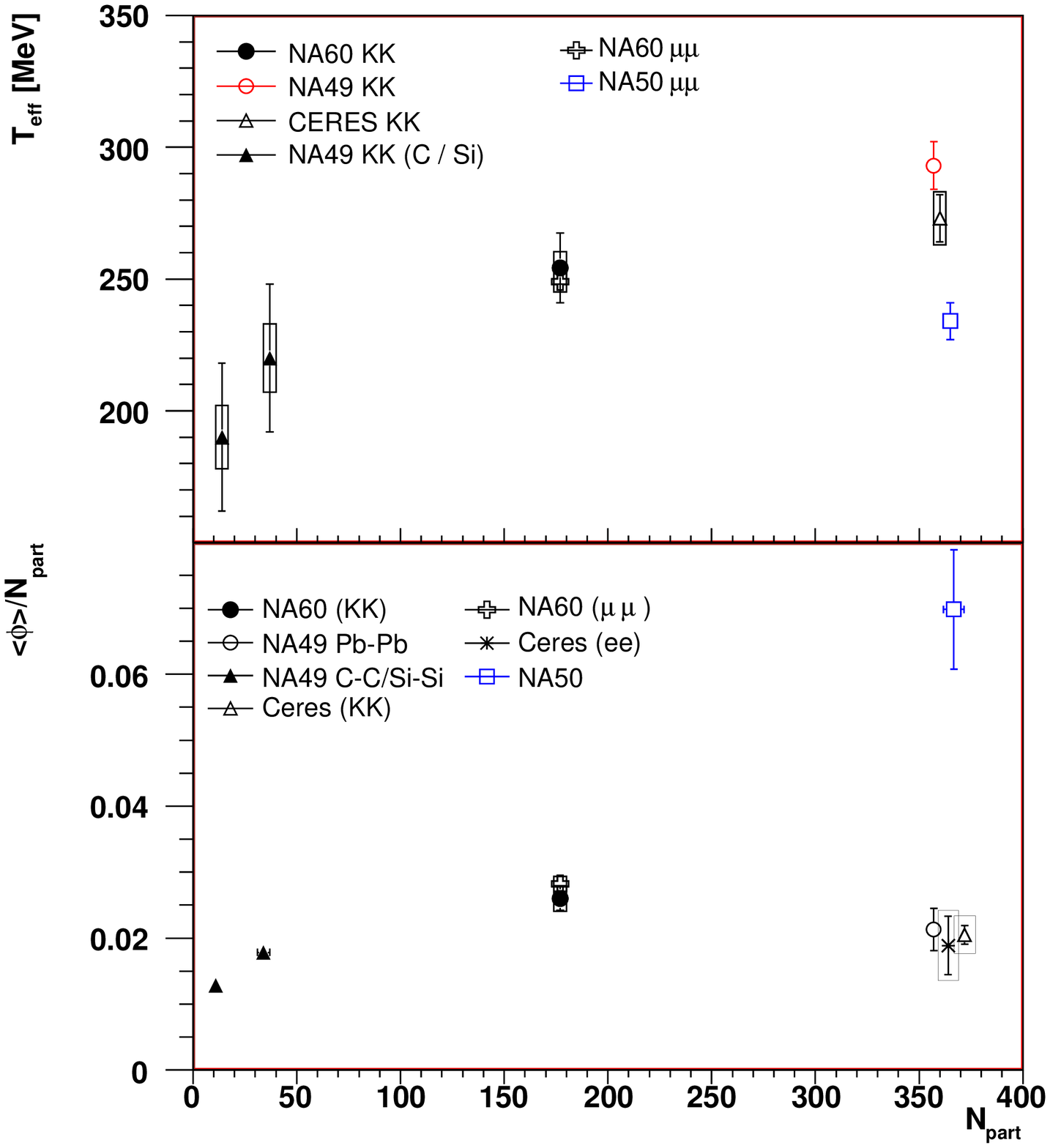}
\caption[]{Inverse slope (top) and enhancement factor (bottom) 
as as function of $N_{part}$.}
\label{fig:collSyst}
\end{minipage}
}
\end{figure}

The invariant mass distribution after background subtraction integrated over 
centrality is plotted in fig.~\ref{fig:mass}. A clear $\phi$ peak on top of a residual 
background can be seen. In order to extract the number of $\phi$ mesons, the mass 
spectrum is fitted with a function that describes the shape of the $\phi$ 
peak superimposed on an empirical description of the residual background. 
The shape of the $\phi$ peak is determined fitting the mass distribution obtained 
from a Monte Carlo simulation with a gaussian superimposed to an empirical 
function that takes into account the high mass tail.  
\\
Several functions and fit mass ranges have been tested to describe 
the residual background. In order to check that a given function 
does not produce artificial bumps in a certain fit range, the fit procedure 
has been applied to the like-sign invariant mass spectra in several 
$p_T$ intervals, where no signal is expected. The functions and fit mass range used 
do not give a fake signal. 
It has to be noticed that if the $\phi$ peak position and width are left 
as free parameters of the fit, the corresponding
values, integrated in centrality, are $m_\phi = 1019.5 \pm 0.3$~MeV/$c^2$ and 
$\sigma_m=7.8 \pm 0.3$~MeV/$c^2$, in good agreement with the simulations.   
The results shown in the following are obtained fixing $m_\phi$ and $\sigma_m$ 
according to the Monte Carlo values. \\
In order to extract the $m_T$ distributions, the fit to the invariant mass 
spectra was performed in several $p_T$ intervals. 
Due to statistics limitations, a reliable $m_T$ distribution could be 
extracted only for central and semi-central collisions. 
Results are reported in fig.~\ref{fig:mtspectra}.
The distributions in the $\phi\rightarrow \mu \mu$ channel in the same 
centrality intervals are reported for comparison. 
The $m_T$ spectra are fitted with the function
$1/m_T dN/dm_T \propto e^{-m_T/T}$. The extracted $\chi^2/ndf$ is
$\sim 1$ in both cases. The inverse slopes measured in semicentral
and central collisions are $253 \pm 11 \pm 5$~MeV and 
$254 \pm 13 \pm 6$~MeV respectively. 
The systematic error is dominated by the choice of 
the function used for the background. 
Results are in good agreement with the ones obtained in dimuons in the 
full $p_T$ range ($250 \pm 2 \pm 3$~MeV 
and $249 \pm 3 \pm 4$~MeV for semicentral and central collisions). 
Since in presence of radial flow the T value may depend on the 
$p_T$ range, being higher at low transverse momenta,
the fit to the dimuon spectra was 
restricted to the range $p_T>0.9$~GeV/$c$, giving $T=252\pm 4$ 
and $247 \pm 3$~MeV for semicentral and central collisions, still in agreement
with the measurement in kaons.\\ 
The raw $\phi$ multiplicity was determined fitting the mass spectra for 
$p_T>0.9$~GeV/$c$ and dividing the number of $\phi$ mesons obtained by the 
total number of events selected for this analysis. This value was then corrected 
for the branching ratio in kaon pairs and 
for the acceptance in 4$\pi$, evaluated through a Monte Carlo simulation. 
Results are plotted in fig.~\ref{fig:yield}. The main contributions to the 
systematic errors are given by the uncertainty in the choice of the fit function 
for the residual background component in the fit of the invariant mass spectra
and the variation of the acceptance caused by the error in the inverse slope. \\
In the same figure, the corresponding results for dimuons are shown. 
In can be seen that the yields in the hadronic and dileptonic channels 
are in agreement within the errors. A ratio between 
the $\phi$ yields in dimuons and in kaons larger than 1.2 is excluded at $95\%$ C.L.\\
%
%
%
%
In order to compare to other collision systems, inverse slope and the enhancement 
factor $\langle \phi \rangle / N_{part}$ are plotted in fig.~\ref{fig:collSyst} as a 
function of the number of participants for central C-C, Si-Si~\cite{na49ccsi}, In-In and Pb-Pb collisions.  
The inverse slope shows an initial fast increase at low $N_{part}$ values, that becomes 
less pronounced going towards higher $N_{part}$.
Lower values are observed by NA50 as compared both to the CERES and NA49 
measurements in Pb-Pb and to the NA60 In-In points in the hadronic and dileptonic 
channels. As stated above, in presence of radial flow T depends on $p_T$. 
As a consequence, the NA49 and  NA50 values can not be directly
compared, since the former is dominated by low $p_T$ ($<1.6$~GeV/$c$) while the 
latter is limited by acceptance to $p_T>1.1$~GeV/$c$. 
However, the NA60 analysis in dimuons, performed
in both $p_T$ ranges, showed a difference of about 15~MeV, which can be probably 
ascribed to a modest radial flow. Radial flow alone can not justify the large difference 
(about 70~MeV) seen between kaon and dimuon results in central Pb-Pb collisions.
A further flattening caused by kaon rescattering and absorption may lead to larger T 
values in the NA49 data. However, this effect is not seen when comparing the In-In 
results in the hadronic and dileptonic channels.\\
Concerning the enhancement factor, an unambiguous comparison to NA50 can 
not be performed in full phase space, since the NA50 acceptance is limited to 
$p_T>1.1$~GeV/$c$ and there is no consensus on the value of the inverse slope 
parameter. In fig.~\ref{fig:collSyst} the NA50 result is extrapolated to full phase space
using a T value of 220~MeV, according to the NA50 measurement in peripheral collisions. 
Even assuming as an extreme case a T value of 300~MeV, the NA50 enhancement factor
would exceed by a factor of $\sim 2$ the central Pb-Pb values 
obtained by NA49 and CERES. Moreover, the CERES measurement in dielectrons 
and kaon pairs are in agreement within the errors, excluding a yield in dileptons exceeding
the one in kaons by more than 60$\%$.  At present,
it is difficult to reconcile all of the observation into a coherent
picture, albeit there is some hint for a possible physics mechanism
leading to a difference in the two channels in Pb-Pb collisions, while in In-In collisions 
no remarkable difference is observed.


\end{document}